\documentstyle[12pt]{article}
\newcommand{\draft}{
        \renewcommand{\baselinestretch}{1.0}%
        \small\normalsize%
}

\draft
\begin{document}
\title{\bf The Strong Magnetic Field Decay and Evolution of Radio Pulsars 
on the P--\.{P} Diagram}
\author{Oktay H. Guseinov$\sp{1,2}$
\thanks{e-mail:huseyin@gursey.gov.tr},
A\c{s}k\i n Ankay$\sp1$
\thanks{e-mail:askin@gursey.gov.tr}, \\
Sevin\c{c} O. Tagieva$\sp3$
\thanks{email:physic@lan.ab.az}, 
\\ \\
{$\sp1$T\"{U}B\.{I}TAK Feza G\"{u}rsey Institute} \\
{81220 \c{C}engelk\"{o}y, \.{I}stanbul, Turkey} \\
{$\sp2$Akdeniz University, Physics Department,} \\ 
{Antalya, Turkey} \\
{$\sp3$Academy of Science, Physics Institute,} \\
{Baku 370143, Azerbaijan Republic} \\
}

\date{}
\maketitle
\begin{abstract}
\noindent
In this work we have analysed various data on radio pulsars and we have 
shown that magnetic field decay of a factor about 10-20 is necessary to 
explain their evolution, in particular to remove the discrepancy between 
the characteristic and the real ages. The character of the field decay is 
exponential with a characteristic time of about 3$\times$10$^6$ yr. 
Observational data on single X-ray pulsars which radiate due to cooling 
also support this result.  

\end{abstract}

\section{Introduction}
Spin period (P) and time derivative of the spin period (\.{P}) of 
radio pulsars (PSRs) 
are known highly precisely and using these quantities one can 
precisely estimate: 1) rate of rotational energy loss (\.{E}), 2) 
effective (overestimated) or real value of the perpendicular component 
of the magnetic field (B) depending on the presence or absence of 
additional torques other than the magnetodipole radiation 
torque, 3) the value of characteristic time ($\tau$) which is the 
characteristic age if the values of 
the braking index n are known along the evolutionary tracks. Since the 
change in n with respect to time is not known, $\tau$ has no physical 
meaning and it is not related to the age of the PSR especially for the 
case of older PSRs. Such definitions are necessary under the existence 
of strong magnetic field decay throughout the evolution of PSRs. 
Please note that as we can estimate only the component of the magnetic 
field (but not the magnetic field itself) perpendicular to the spin axis 
using P and \.{P} values, by saying 'magnetic field decay' we mean a 
decrease in the value of the perpendicular component. 

Although the evolution of PSRs on 
the P-\.{P} diagram has been studied since 1970s, there are still some 
uncertainties and open questions about the evolution of these objects. 
$^{1-3}$ Below, widely used definitions of $\tau$ and B (i.e. for n=3) 
will be used unless another definition is given. 

On the P-\.{P} diagram, PSRs must move along  
constant magnetic field lines if no additional torque other than 
the magnetodipole radiation torque is present and/or no decay in the 
magnetic field occurs and/or the rotation and the magnetic axes do not 
align in time. Actually, these processes may take place with different 
degrees for different PSRs and this has been discussed since 
seventies. $^{4-10}$ On the other hand, radio luminosity of PSRs may 
decrease several times during the evolution. When PSRs approach 'the 
death belt' (of which the place on the P-\.{P} diagram is not known well) 
the effect of nulling (in small degree), narrowing of the radiation beam, 
and turning-off of PSRs begin to take place. All of these create 
difficulties in determining the evolutionary tracks of PSRs on the 
P-\.{P} diagram. Note that the origin and evolution of radio PSRs seem 
to be known better and investigating this type of pulsar is 
easier compared to other types of pulsar and in particular to other types 
of neutron star. In this work, we try to improve and advance in this very 
important problem. 

\section{What do we see on the P-\.{P} diagram?}
Today, one can estimate PSR distances with errors $\sim$30\% if they 
are located up to about 5-6 kpc from the Sun. $^{11}$ Most 
of the far away PSRs have been discovered in recent surveys 
(southern sky close to the Galactic plane $^{12}$). So, we will limit 
our statistical investigation by examining only the PSRs located at 
d$<$4 kpc from the Sun. In this region, there are about 600 single 
PSRs out of about 1400 PSRs detected up to date in the Galaxy. 

The rate of rotational energy loss of PSRs is 
\begin{equation}
\dot{E} = \frac{32 \pi^4}{3c^3} \frac{R^6 B^2}{P^4}
\end{equation}
$^{10,13}$ where R is the radius and c is the speed of light. As seen 
from expression (1), positions of the PSRs with small values of B and 
large values of P must change very slowly on the P-\.{P} diagram if the 
component of the magnetic field perpendicular to the spin axis does not 
diminish. On the other hand, parameters of young PSRs considerably 
rapidly change compared to PSRs with large characteristic time 
($\tau$=P/(n-1)\.{P}). Therefore, we begin to investigate the single-born 
PSRs with the values \.{P}$>$10$^{-17}$ s/s. 

All the known PSRs with \.{P}$>$10$^{-17}$ s/s are represented in 
Figure 1. As seen from this figure, the direction of the increase in the 
number density of PSRs on the P-\.{P} diagram does not coincide with 
the directions of the lines of constant magnetic field. For 
PSRs with
10$^{12}$$<$B$<$10$^{13}$ G and $\tau$$>$10$^5$-10$^6$ yr, this may be 
related to PSRs entering the 'death-belt'. But these lines also do not 
coincide in the region where $\tau$$<$10$^6$ yr. For the PSRs with 
$\tau$$<$10$^5$ yr the observed values of the braking index (n) are less 
than or approximately equal to 3. $^{14}$ The value of n must be 
greater than 3 during and immediately after the magnetic field decay when 
influence of the activity processes of the PSRs begin to come to an 
end. Therefore, we may say that there are also some observational data 
which confirm the magnetic field decay for the youngest PSRs, a 
natural decay as a result of high temperatures. $^{15}$
On the other hand, such an effect can also be due to a considerable 
decrease in the beaming fraction of PSRs as the value of P increases, 
e.g. a decrease proportional to P$^{-1/2}$. $^{16-18}$ As the number of 
the PSR data is very large, we have constructed a dependence between the 
ratio of the pulse width to the period and the period and also a 
dependence between this ratio and $\tau$. 
We see that the known dependence between the beaming fraction and the 
period remains valid.  

As the PSR searches have mainly taken place in the Galactic plane 
(especially in the last few years) and as the old PSRs can go far away 
from the Galactic plane, the number of young PSRs among the observed 
ones in the surveys must be relatively large. So, the number density of 
PSRs on the P-\.{P} diagram must be shifted to large values of $\tau$ 
if we consider only the PSRs close to the Sun. In Figure 2 we have 
plotted only the PSRs with values of projected distances on the plane 
of the Galaxy d$\times$Cos b $\le$ 3 kpc. The distance values were taken 
from Guseinov et al. $^{11}$ and these 
values are similar to the ones given in ATNF catalogue $^{12}$. As seen 
from a comparison of this figure with Figure 1, there is 
an increase in the number density distribution of PSRs with large 
values of $\tau$. (In Figure 2, the number density of PSRs with 
$\tau$$<$10$^6$ yr and with large values of B is considerably less.) But 
the increase is not so much as must be under 
linear dependence between the number density and $\tau$. And this must 
not only connected to low sensitivity of the surveys for large Galactic 
latitude values at which most of the PSRs with large values of $\tau$ 
must be located.

Is this mainly related to a considerable decrease in the luminosity with 
the increase in characteristic time? It is true that this can not be a 
strong effect, but it is necessary to consider the influence of it.  
In Figure 3, the 30 PSRs which have the largest luminosities at 
1400 MHz (L$_{1400}$$\ge$ 9.0 mJy kpc$^2$) are denoted with the symbol 
'X' and the 50 PSRs with the smallest values of luminosity 
(L$_{1400}$$\le$ 1.58 mJy kpc$^2$) are represented with 'circles'. As 
seen from the figure, the average luminosity of PSRs may decrease a 
factor of only a few times during the evolution while the differences 
between the initial luminosity values of PSRs are about 4-5 orders of 
magnitude. $^{10,19}$ We 
have constructed also a similar figure for PSRs with d$\times$Cos b 
$\le$ 3 kpc to check this result. Again, there is no considerable 
difference in the average luminosity values of young and old PSRs. 
Therefore, diminishing of the luminosity can not have an important role 
in the 'death' of PSRs with \.{E}$>$10$^{31}$ erg/s. 
 
Above, we see that the number of PSRs with 
10$^{11}$$\le$B$\le$10$^{13}$ G does not increase proportionally to the 
value of $\tau$. In order to make a better analysis, let us consider the 
P-\.{P} diagram of the PSRs with d$\times$Cos b $\le$3 kpc displayed 
in Figure 2. The numbers of PSRs with 10$^{11}$$\le$B$\le$10$^{13}$ G 
and with characteristic times $\le$10$^5$, $\le$10$^6$, $\le$10$^7$ and 
$\le$10$^8$ yr are written in the caption of Figure 2.

We have tested the change of the number of PSRs with respect to 
$\tau$ also by adopting a PSR sample in a cylindrical volume with 
d$\times$Cos b $\le$2 kpc, because 
we want to see the influence of PSRs going out from and coming in to 
the volumes under consideration. We see that the ratio of the number of 
PSRs with respect to $\tau$ changes with $\tau$ practically similarly 
in both cases. This ratio is about 4.5-4.6 when $\tau$ changes from 
10$^5$ to 10$^6$ yr and from 10$^6$ to 10$^7$ yr. But when the value of 
$\tau$ increases from 10$^7$ yr up to 10$^8$ yr, the number ratio of
PSRs is only about 1.7 (in both cases). This slow increase in the 
number of PSRs under the increasing value of $\tau$ is a result of the 
PSRs going deeper through the death-belt where it is not certainly 
known what processes lead to the turn-off of PSRs and the decay of the 
magnetic field. We only know that this effect strongly depends on the 
values of B and P and hence on \.{E}. The rate of 
PSRs' turn-off increases with the increase in the values of B and P.
This can directly be seen from Figure 2. The decrease in the number 
of some of these PSRs up to $\tau$$\sim$10$^7$ yr may be related to the 
decreasing of the beaming fraction, but the fact that the PSRs with 
$\tau$$>$10$^7$ yr are deficient is weakly dependent on the beaming 
fraction. If we consider only the PSRs with  
10$^{11}$$\le$B$\le$10$^{12}$ G, then the number 
ratio with respect to $\tau$ is $\sim$7.9 and $\sim$2.4 under the 
increase of $\tau$ from 10$^6$ to 10$^7$ yr and from 10$^7$ to 10$^8$ yr, 
respectively, for both considered volumes. The considerably large number 
ratio in this 
B interval may be the result of strong magnetic field decay under which 
PSRs in the interval 10$^{12}$$\le$B$\le$10$^{13}$ G penetrate into the 
interval 10$^{11}$$\le$B$\le$10$^{12}$ G. In the latter magnetic field 
interval PSRs can not reach to large values of period as they do under 
large values of B, because the death-belt is parallel to the constant 
\.{E} lines. As a result of the diminishing of the perpendicular component 
of the magnetic field, PSRs penetrate through the death-belt and turn-off 
in considerably small time intervals. Thus, the increase in the 
difference between the real age and the characteristic time under the 
increasing of the value of $\tau$ can be understood. 

\section{A direct evidence for the strong magnetic field decay}
In the previous section, we see that the number density of PSRs increases 
in the direction of increasing value of $\tau$, but not as a linear 
dependence. On the other hand, this increase depends on the depth of 
penetration through the death-belt, the value and the decay of the 
magnetic field, and the period values of PSRs. Let us now discuss this 
problem using another approach.

The average distance of PSRs from the plane of the Galaxy is 
\begin{equation}
\overline{\mid z \mid} = \overline{\mid V_z \mid} \times t
\end{equation}
where V$_z$ is the value of the z-component of the space velocity 
and t is the real age of PSR. As
\begin{equation}
t = \frac{P}{(n-1)\dot{P}} (1-(\frac{P_0}{P})^{n-1})
\end{equation} 
value of $\tau$ is equal to or linearly depends on t for small values of 
initial spin periods (P$_0$$<<$P) and 
\begin{equation}
\overline{\mid z \mid} \sim \tau.
\end{equation}
In Figure 4, $\mid$z$\mid$ versus $\tau$ distribution of the PSRs with 
d$\times$Cos b $\le$ 3 kpc and \.{P}$>$10$^{-17}$ s/s is represented. 
Effect of the radio luminosity 
and the sensitivity of the latest PSR surveys may have some influence on 
this dependence. In order to exclude (or diminish) this effect, we have 
used only the PSRs which were observed at 400 MHz and which have 
log L$_{400}$ = log (F$_{400}$$\times$d$_{kpc}^2$) $\ge$ 1. As seen from 
Figure 4, the average value of $\mid$z$\mid$ increases about a factor of 2 
if we compare the PSRs in the intervals 10$^6$ $<$ $\tau$ $<$ 10$^7$ yr 
and 10$^7$ $<$ $\tau$ $<$ 10$^8$ yr (this increase may be up to a factor 
of 3 if we take into account the decrease in PSR luminosity and comparably 
worse search of PSRs at large latitude values). When we compare the PSRs 
in the intervals 10$^7$ $<$ $\tau$ $<$ 10$^8$ yr and 10$^8$ $<$ $\tau$ 
$<$ 10$^9$ yr we see that the average value of $\mid$z$\mid$ practically 
does not increase beyond $\tau$$\sim$10$^7$ yr. This strongly 
contradicts the existence of a linear dependence between the real and the 
characteristic ages for PSRs with $\tau$$>$10$^6$ yr and the evolution of 
PSRs along B=const lines if the component of the magnetic field 
perpendicular to the rotation axis does not decrease. In principle this 
was known many years ago but on the basis of small statistical data and 
with larger uncertainties in the distance values of PSRs. The differences 
between the characteristic times and the kinematic ages do not depend on 
the beaming fraction nor on the effects which may influence the character 
of the number density change under the increasing of $\tau$. 

We have also constructed  the dependence of $\overline{\mid z \mid}$ on 
$\tau$ for the same volume as in Figure 4 but for the PSRs with 
10$^{11}$ $<$ B $<$ 10$^{12}$ G. The average values of 
$\mid$z$\mid$ in different $\tau$ intervals for this sample are 
practically the same as in Figure 4. It is especially necessary to note 
that kinematic ages of the PSRs with log $\tau$$\ge$8.5 are about 2 
orders of magnitude smaller than the $\tau$ values as seen from Figure 4. 
As real ages of the PSRs with log $\tau$$\ge$7.5 are considerably smaller 
than the $\tau$ values their location on this part of the P-\.{P} diagram 
shows that the component of the magnetic field perpendicular to the 
rotation axis diminishes more than one order of magnitude (see Fig. 2).

From Figure 4, not only the absence of a linear dependence between 
the real ages and the characteristic times but also the absence of a 
linear dependence of the number density of PSRs on $\tau$ is seen, 
similar to Figure 2. These effects may strongly be the result of also 
the influence of the decrease in the beaming angles (and in accordance 
the beaming fraction which is related to it) as t and $\tau$ increase. 
Of course it is in more degree a result of the decrease in the magnetic 
field and the magnetic polar areas. 

In Figure 2, the PSRs which have also been detected in X-rays are 
displayed with sign 'X'. The X-ray luminosity strongly depends on the 
value of \.{E} $^{20-22}$ and the \.{E} values of the PSRs in Figure 
2 are in a wide interval from $\sim$10$^{37}$ erg/s down to 
$\sim$10$^{29}$ erg/s. Therefore, the number density of the PSRs with 
small L$_x$ values is very high, though this is not seen directly from 
Figure 2 which represents a considerably large volume. If we take into 
account these facts, then from the locations of these objects on the 
P-\.{P} diagram we can be sure about the decay in the magnetic field (by 
diminishing of the magnetic field component which is perpendicular to the 
rotational axis) of a factor about 3-4 until $\tau$$\sim$10$^6$ yr. 

The large birth rate of PSRs with B$>$10$^{13}$ G, which later drop down 
to 10$^{11}$$<$B$\le$10$^{12}$ G interval, supports their considerably 
large number density at large values of $\tau$. Although the 
magnetic field of these PSRs decreases, in general they remain in the 
considered interval of the magnetic field. The braking index can be also 
less than 3 at some points during the evolution up to $\tau$$\sim$10$^6$ 
yr, but the dominant factor must be the field decay and so the braking 
index must be greater than 3 throughout the whole evolutionary track.    

Let us now discuss the character of the magnetic field decay in relation 
to the observational data. Above, from the analysis of the observational 
data, we have seen that the component of the magnetic field perpendicular 
to the rotational axis must decrease and the character of this decrease 
must roughly be a simple exponential law. In this case, the characteristic 
time ($\tau$=P/2\.{P} for n=3) of PSRs depends on both the characteristic 
time of the field decay ($\tau$$_d$) and the real age (t): 
\begin{equation}
\tau = \frac{\tau_d}{2} e^{2t/\tau_d}
\end{equation}
if t$>>$$\tau$$_d$. $^4$ If we adopt 
$\tau$$_d$=3$\times$10$^6$ yr, then the evolutionary tracks of PSRs 
approximately lead to the actual distribution of PSRs on the P-\.{P} 
diagram. In order to estimate the values of P and \.{P} we can use the 
equation 
\begin{equation}
P \dot{P} = \frac{B_0^2}{10^{39}} e^{-2t/\tau_d}.
\end{equation}
In Table 1, t, $\tau$, P and \.{P} values (found from eqn.(4) and eqn.(5)) 
for 2 different values of B$_0$ are represented. The positions 
of PSRs (corresponding to t=6$\times$10$^6$ yr and 9$\times$10$^6$ yr)
with initial magnetic field values 10$^{12}$ 
G and 10$^{13}$ G under the exponential magnetic field decay with a 
characteristic time of 3$\times$10$^6$ yr are displayed in Figure 2.

\section{Discussion and Conclusions}
As known, the X-ray radiation depend 
very strongly on the value of rate of rotational energy loss (\.{E}) of 
PSRs. $^{20-22}$ Moreover, the acceleration of particles mainly take 
place in the field of magnetodipole radiation wave. $^{23}$ Therefore, 
the 'death-belt', as commonly adopted, passes parallel 
to the constant \.{E} lines. PSR radio radiation may begin to turn off 
practically for $\tau$$\sim$10$^6$-10$^8$ yr depending on the 
magnetic field value (correspondingly 10$^{13}$-10$^{11}$ G), in other 
words, beginning from \.{E}$\sim$10$^{32}$-10$^{31}$ erg/s. Processes for 
the turn-off become very strong as \.{E} decreases. In spite of these 
processes, the radio radiation of PSRs practically do not decrease 
till the turn-off or we can say that the decrease in their radio radiation 
is not important in our investigation. The evolutionary tracks of PSRs on 
the P-\.{P} diagram 
and the places of the turn-off strongly depend on the magnetic field decay 
which has roughly an exponential character. As follows from our analysis 
of the data, the characteristic time of this decay is 
about 3$\times$10$^6$ yr. Following this discussion we understand 
that the main and dominant reason for the considerable deviation of the 
direction of the increase in the number density of PSRs (direction 
of evolution) from the constant B lines is the magnetic field decay and 
the turning off of PSRs under the same values of \.{E}. 

The problem under discussion is the oldest one in PSR astronomy. As we 
have seen in the sections above, the main reason to suggest a magnetic 
field decay more than one order of magnitude up to $\sim$10$^7$ yr 
is that the kinematic ages of PSRs practically 
do not increase for $\tau$$\ge$3$\times$10$^7$ yr (see Figure 4). 
The strong decay of the magnetic field also comes from the distribution of 
single X-ray PSRs including the PSRs which radiate due to cooling. In 
principle, we can explain the direction of the increase in the number 
density of PSRs on the P-\.{P} diagram and the absence of a correlation 
between the number of PSRs and $\tau$ based on decreasing of the 
luminosity and the turn-off. But these ways are too artificial 
compared to explaining the direction of the increase in the number density 
by magnetic field decay.

The idea of exponential decay of the magnetic field of PSRs was more 
popular in the past, before the observation of millisecond PSRs with 
weak magnetic fields (i.e. B$<$10$^{10}$ G). After estimating the magnetic 
field of neutron stars in X-ray binary systems, the idea of magnetic field 
decay without accretion was rejected. In this work we see from the 
analysis of the observational data that the component of the magnetic 
field perpendicular to the spin axis decreases significantly. The 
decay must 
weaken after t$\sim$10$^7$ yr. We think that Ruderman $^3$ may be right 
not only about the character of the magnetic field decay but also about 
the increasing or the invariability of the magnetic field when PSRs are 
very young. There exist many complex problems about the evolution of  
magnetic field of neutron stars, which date back to the seventies. 
$^{24,25}$ The magnetic field 
decay is necessary to explain the evolution of PSRs based on the  
actual/available observational data and especially to get rid of the 
discrepancy between $\tau$ and t.

\clearpage
\begin{flushleft}
\begin{tabular}{ccccc}
\multicolumn{5}{l}{\bf Table 1 - The values of t, $\tau$, P and \.{P} 
found from} \\ 
\multicolumn{5}{l}{\bf equations (4) and (5) for 2 different values of 
B$_0$} \\ 
\multicolumn{5}{l}{\bf ($\tau$$_d$=3$\times$10$^6$ yr).} \\ \hline 
B$_0$ & t & $\tau$ & P & \.{P} \\
(10$^{12}$ G) & (10$^6$ yr) & (10$^6$ yr) & (s) & (10$^{-18}$ s/s) \\ 
\hline
10 & 6 & 80 & 3.0 & 610 \\
& 9 & 583 & 3.1 & 84 \\ \hline
1 & 6 & 80 & 0.3 & 60 \\
& 9 & 583 & 0.31 & 8.38 \\ \hline

\end{tabular}
\end{flushleft}

\clearpage
{\bf Figure Captions} \\
{\bf Figure 1:} P-\.{P} diagram of all PSRs with \.{P}$>$10$^{-17}$ s/s 
in the ATNF Pulsar Catalogue. \\
{\bf Figure 2:} P-\.{P} diagram of all PSRs with d$\times$Cos b 
$\le$ 3 kpc. The PSRs which have also been detected in X-rays are 
denoted with 'X'. There are 8 PSRs with $\tau$$\le$10$^5$ yr, 41 
PSRs with $\tau$$\le$10$^6$ yr, 181 PSRs with $\tau$$\le$10$^7$ yr, 
and 299 PSRs with $\tau$$\le$10$^8$ yr. The 'black squares' show the 
positions of PSRs under exponential decay of the perpendicular component 
of the magnetic field: the two squares on the right are for the case of 
B$_0$=10$^{13}$ G (for the upper one t=6$\times$10$^6$ yr and for the 
lower one t=9$\times$10$^6$ yr) and the ones on the left are for 
B$_0$=10$^{12}$ G (again for the upper one t=6$\times$10$^6$ yr and for 
the lower one t=9$\times$10$^6$ yr). \\ 
{\bf Figure 3:} P-\.{P} diagram of PSRs with d$\times$Cos b $\le$ 2 kpc.
The 30 PSRs which have the highest L$_{1400}$ values are displayed with
'X'. The 50 PSRs which have the lowest L$_{1400}$ values are shown as
'circles'. The other PSRs are represented with 'plus' signs. \\
{\bf Figure 4:} $\mid$z$\mid$ versus log $\tau$ diagram of the PSRs 
which have \.{P}$>$10$^{-17}$ s/s, d$\times$Cos b $\le$ 3 kpc and log 
L$_{400}$ $\ge$ 1.\\

\clearpage
\begin{figure}[t]
\vspace{3cm}
\includegraphics{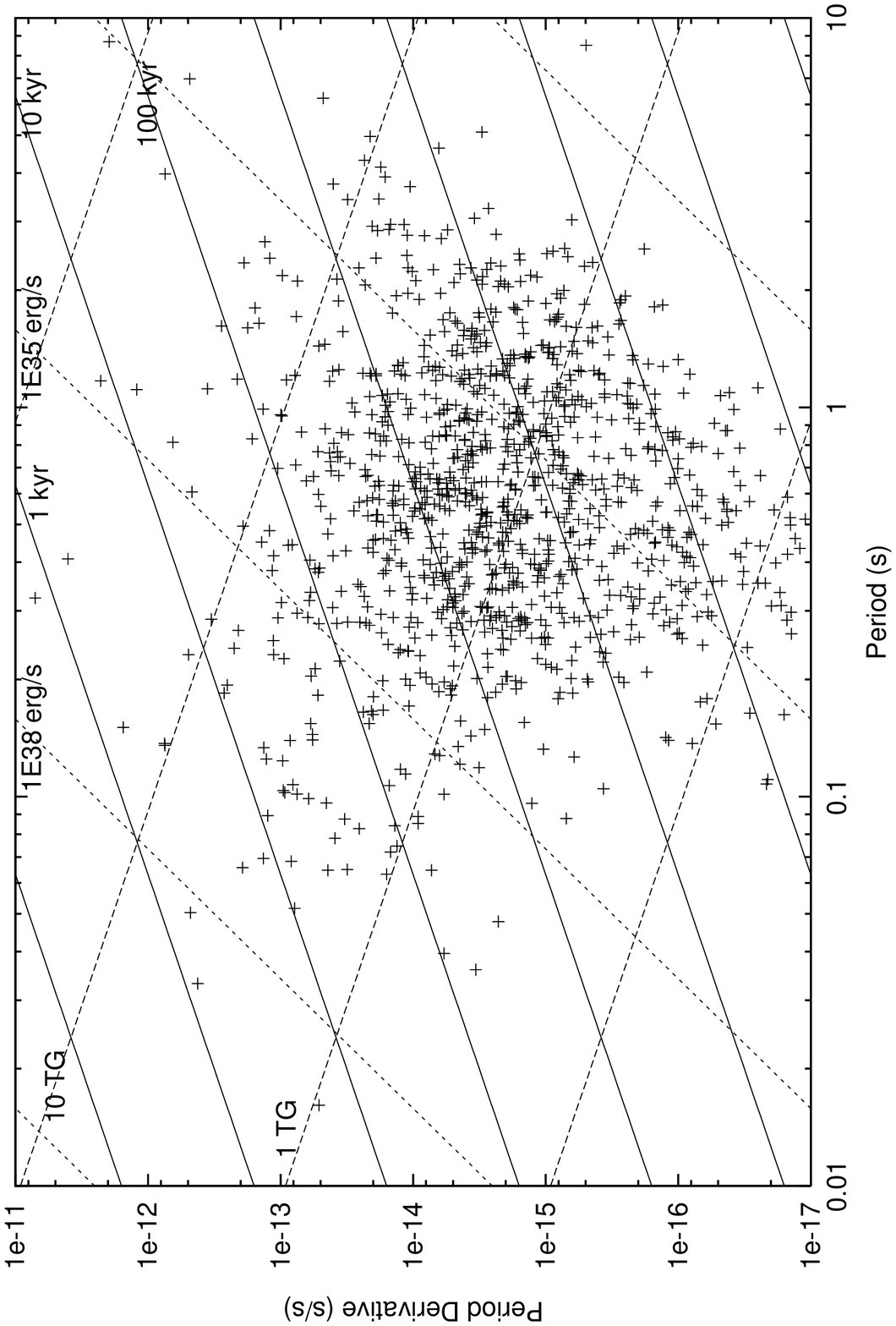}
\end{figure}

\clearpage
\begin{figure}[t]
\vspace{3cm}
\includegraphics{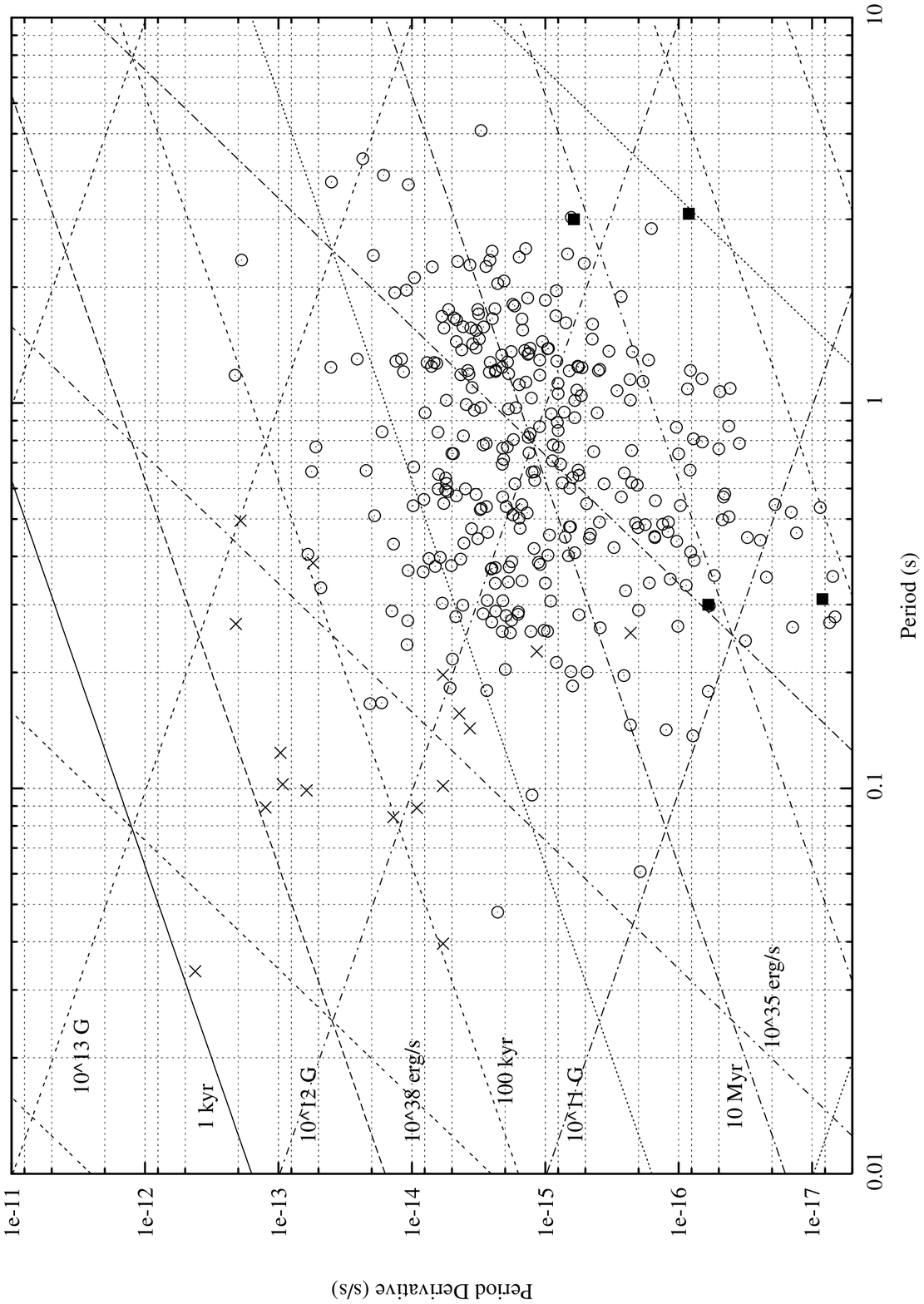}
\end{figure}

\clearpage
\begin{figure}[t]
\vspace{3cm}
\includegraphics{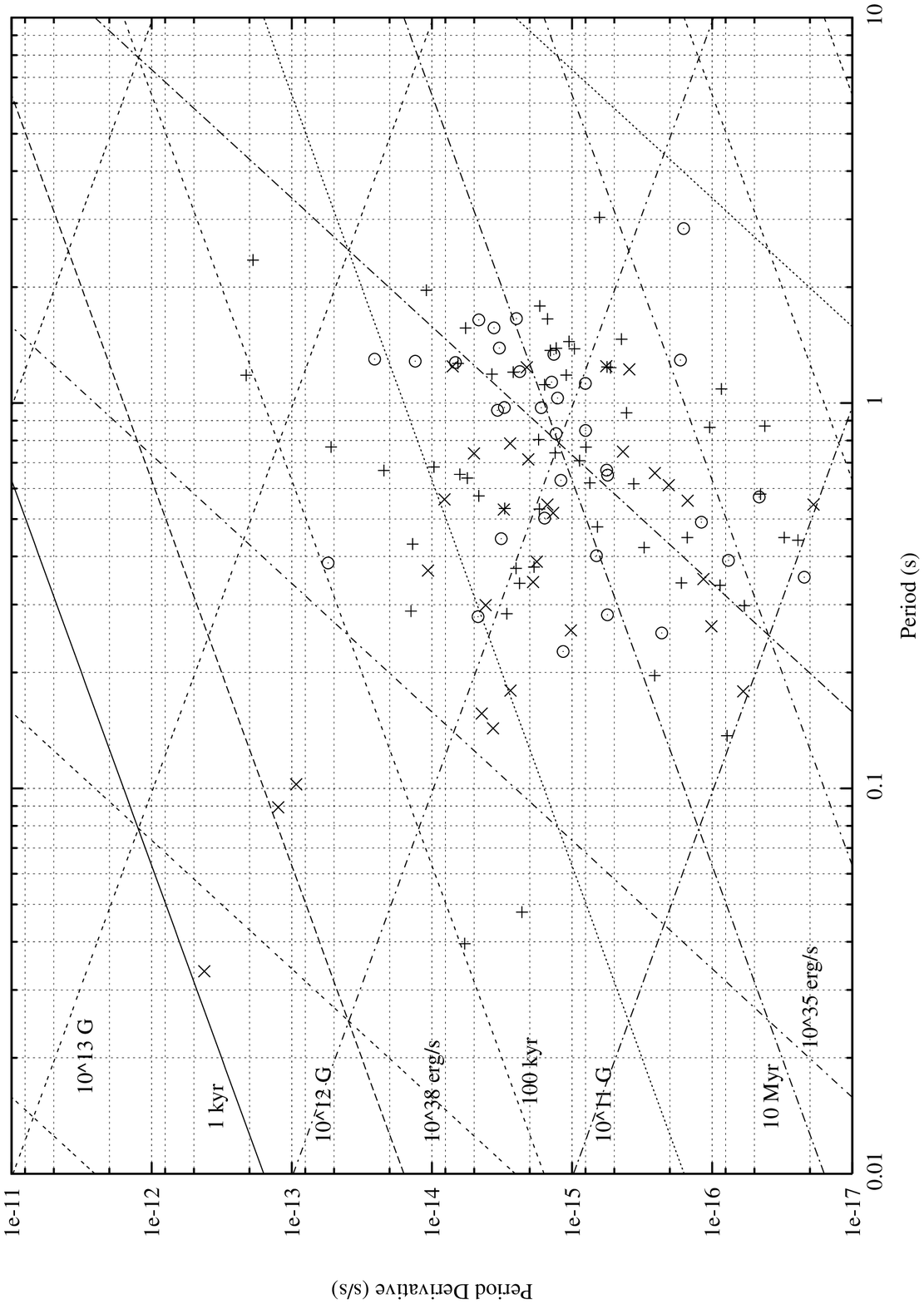}
\end{figure}

\clearpage
\begin{figure}[t]
\vspace{3cm}
\includegraphics{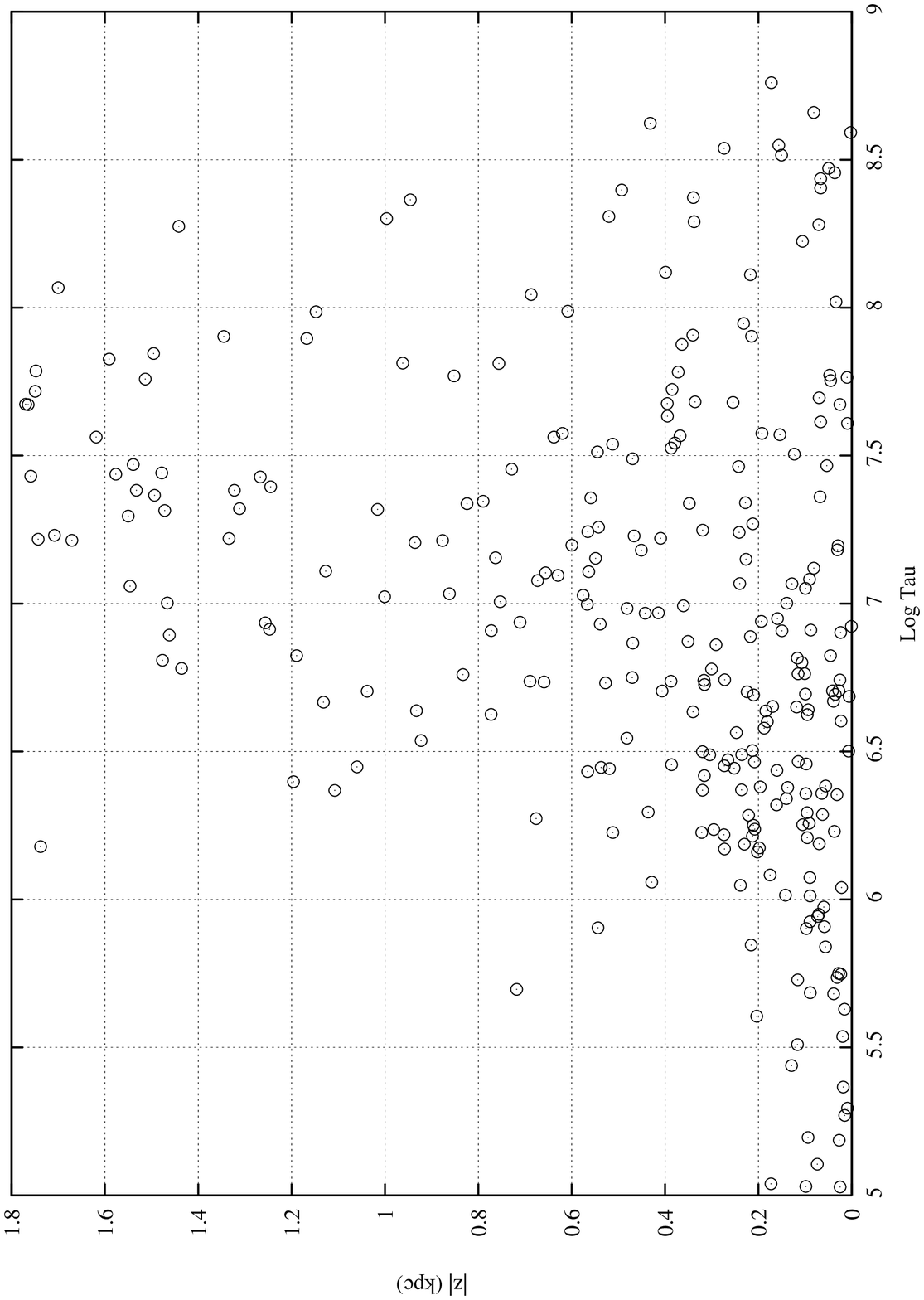}
\end{figure}

\end{document}